\documentstyle[epsfig]{aipproc}

\def\etalk{{\it et al., }}

\newcommand{\xpom}{x_{_{\!I\!\!P}}}

\newcommand{\lapprox}{\stackrel{<}{_{\sim}}}
\newcommand{\pom}{\rm I\!P}

\newcommand{\pt}{p_{_T}}
\newcommand{\mx}{M_{_X}}
\newcommand{\my}{M_{_Y}}
\newcommand{\ftwod}{F_2^{D(3)}(\beta,Q^2,\xpom)}
\newcommand{\alphapom}{\alpha_{_{\rm I\!P}}}
\newcommand{\alphareg}{\alpha_{_{\rm I\!R}}}

\newcommand{\av}[1]{\mbox{$ \langle #1 \rangle $}}
\newcommand{\scaption}[1]{\caption{\protect{\footnotesize  #1}}}

\begin{document}
\title{Colour-Singlet Exchange in {\boldmath $ep$} 
Interactions\footnote{Summary talk from the diffractive sessions of the 1997
DIS workshop, Chicago.}}

\author{P. R. Newman\thanks{Supported by the UK Particle Physics and
Astronomy Research Council.}}
\address{
School of Physics and Astronomy, University of Birmingham B15 2TT, UK}
\maketitle

\begin{abstract}
  \noindent Results presented at the DIS97 workshop by the H1, ZEUS and E665 
collaborations on processes yielding large rapidity gaps and energetic
leading baryons are reviewed. A consistent picture begins to emerge in which 
diffractive processes dominate when the fractional longitudinal momentum loss
at the baryon vertex $\xpom$ is small, with substantial
contributions from other processes as $\xpom$ increases. The diffractive
mechanism in the deep-inelastic regime is found, both from inclusive
measurements and final state studies, to involve the exchange
of a gluon carrying a large fraction of the exchange momentum. Vector meson
results show the transition from soft to hard production mechanisms with
increasing precision.
\end{abstract}

\section{Introduction}
\label{prn:intro}

This contribution corresponds to the summary talk given at DIS97
on diffractive $ep$ scattering. It is written as a mainly qualitative 
summary of the results and their interpretations as presented at the workshop. 
It is best read in conjunction 
with\cite{prn:soper}, which contains the other half of the summary from
the diffractive sessions.

The sessions were concerned with the
phenomenological understanding of semi-inclusive experimental 
data that can be summarised by the two closely related diagrams shown in 
figure~\ref{prn2:diagrams}. In the first type of analysis 
(figure~\ref{prn2:diagrams}a), events containing large gaps in the rapidity
distributions of final state hadrons are studied. The hadronic final state is 
divided into two systems $X$ and $Y$, such that
a colour singlet exchange coupling to the $\gamma - X$ and $p - Y$ vertices
can be defined. Most of the results presented were concerned with the case
where the system $Y$ is dominantly a proton. 
Where the system $X$ is a bound state vector meson, the
process is considered elastic at the photon vertex. Dissociation processes
correspond to the cases where one or both of the invariant masses $\mx$ and 
$\my$ of the two final state systems is large. The processes studied cannot
automatically be considered to be diffractive in the sense of the exchange
of the leading vacuum singularity. It is also important to  
understand the role played by non-diffractive processes in the
hadron level cross sections measured.

In addition to the conventional kinematic variables $x$, $Q^2$, $y$ and $W$
used to discuss 
inclusive $ep$ interactions, the following variables are also used here:
\begin{eqnarray}
t = (P - Y)^2 \hspace{1.5cm} \xpom = \frac{q.(P - Y)}{q.P} \hspace{1.5cm}
 \beta = \frac{Q^2}{2 q.(P - Y)}
\label{prn:kine}
\end{eqnarray}
where $q$, $P$ and $Y$ are respectively the four-vectors of the 
incoming photon, incoming proton and outgoing system $Y$ in the framework of
figure~\ref{prn2:diagrams}a. The variable $t$ is the squared 
four-momentum transferred at the proton vertex, $\xpom$ is the fraction of 
the proton beam momentum transferred to the longitudinal momentum of the 
colour singlet exchange and $\beta = x / \xpom$ 
(used only in the DIS regime)
is the fraction of the colour singlet 
exchange 4-momentum that is carried by the quark coupling to the photon.
In the large rapidity gap analyses presented, 
the system $Y$ generally passes unobserved down the beam-pipe
such that $\my$ is constrained to be small. With the exception of
vector meson production, the value of $t$ is not measured. 
The analyses are performed in the kinematic region in which $\mx$ is well 
reconstructed by direct measurement. The relations 
$\xpom \simeq (\mx^2 + Q^2) / (W^2 + Q^2)$ and 
$\beta \simeq Q^2 / (\mx^2 + Q^2)$ are generally
used to reconstruct the remaining kinematic variables. 

\begin{figure}[h] \unitlength 1mm
 \begin{center}
   \begin{picture}(120,25)
     \put(12,3.75){\epsfig{file=prn.diffbasic.epsf,width=0.225\textwidth}}
     \put(80,2.25){\epsfig{file=prn.leadingbary.epsf,width=0.225\textwidth}}
     \put(0,12){\Large{\bf (a)}}
     \put(70,12){\Large{\bf (b)}}
   \end{picture}
 \end{center}
 \vspace{-0.5cm}
 \scaption {The generic process of study for (a) analysis of large rapidity
gap processes ($ep \rightarrow eXY$) and (b) leading baryon analyses
($ep \rightarrow eXN$ with $N = p$ or $n$).}
 \label{prn2:diagrams}
\end{figure}

In the second type of analysis (figure~\ref{prn2:diagrams}b),
a proton or neutron of energy $E_N$ and small transverse momentum
is measured in detectors very near to
the outgoing proton direction. This approach
has the advantage that the system $Y$ is
constrained to be a single state, but also results in a reduction in statistics
and kinematic range by comparison with rapidity gap analyses.
A semi-inclusive cross section can be defined
differentially in $E_N$ throughout the range $0 < E_N < E_p$
where $E_p$ is the incoming proton beam energy.
Where $E_N \sim E_p$, the process is
essentially that of figure~\ref{prn2:diagrams}a in the limit in which the
system $Y$ is a nucleon. All of the kinematic variables defined 
above apply equally
well in both approaches. For the leading baryon
measurements, $\xpom$ is reconstructed using
$\xpom \simeq 1 - E_N / E_p$.

\section{Semi-Inclusive Cross Section Measurements at low {\boldmath $\xpom$}}
\label{prn:incl}

Measurements were presented by both H1 and ZEUS of the semi-inclusive DIS
interaction $ep \rightarrow eXY$ and the corresponding photoproduction process
$\gamma p \rightarrow XY$. The mass of the system $Y$ is constrained to be as 
small as possible such that the
data samples are dominated by the single dissociation process in which 
$Y$ is a single proton. 

In the H1 case, the DIS results are measured using the rapidity gap method 
in the kinematic range $\my < 1.6 \ {\rm GeV}$ and 
$|t| < 1.0 \ {\rm GeV^2}$\cite{prn:dirkmann}. The data are presented
in terms of a three dimensional structure function $\ftwod$. This
is essentially the inclusive structure function $F_2$ differential in $\xpom$
and integrated over the $\my$ and $t$ ranges above;
\begin{eqnarray}
  \frac{{\rm d}^3 \sigma_{e p \rightarrow e X Y}}{{\rm d}\xpom \,{\rm d}\beta\
,{\rm d}Q^2} 
= \frac{4\pi\alpha^2}{\beta Q^4}(1-y+\frac{y^2}{2})F_2^{D(3)}(\beta,Q^2,\xpom).
\end{eqnarray}
A Regge approach\footnote{The description of diffractive DIS, almost by 
definition, requires a mixture of perturbative and non-perturbative 
physics\cite{prn:buchmuller}. A number of different approaches are taken to 
disentangle the hard from the soft aspects. In this contribution, Regge
language is used to discuss the proton vertex, with the hard interaction
viewed as DIS from a distinct set of parton distributions for the exchanged
objects. Discussions of other related and unrelated
models may be found in\cite{prn:soper,prn:buchmuller}.} 
is taken to parameterise 
the $\xpom$ dependence such that $F_2^{D(3)}$ 
decomposes into contributions from the exchange of different trajectories 
$\alpha_i(t) = \alpha_i (0) + \alpha_i^{\prime} t$ according to
\begin{eqnarray}
  \ftwod = \sum_i \int_{-1 \ {\rm GeV^2}}^{t_{\rm min}} {\rm d} t \ 
e^{b_{_0}^i t} \left( \frac{1}{\xpom} \right)^{2 \alpha_i(t) - 1} 
A^i (\beta,Q^2)
  \label{prn:f2dfac}
\end{eqnarray}
where $A^i (\beta,Q^2)$ is proportional to the 
structure function of the exchange $i$ and
$b_{_0}^i$ describes the $t$ dependences of the couplings of the exchange $i$
to the photon and the proton.
For each separate exchange, the $\xpom$ dependence then factorises
from the $\beta$ and $Q^2$ dependence. 

\begin{figure}[h] \unitlength 1mm
 \begin{center}
   \begin{picture}(160,45)
     \put(0,-5){\epsfig{file=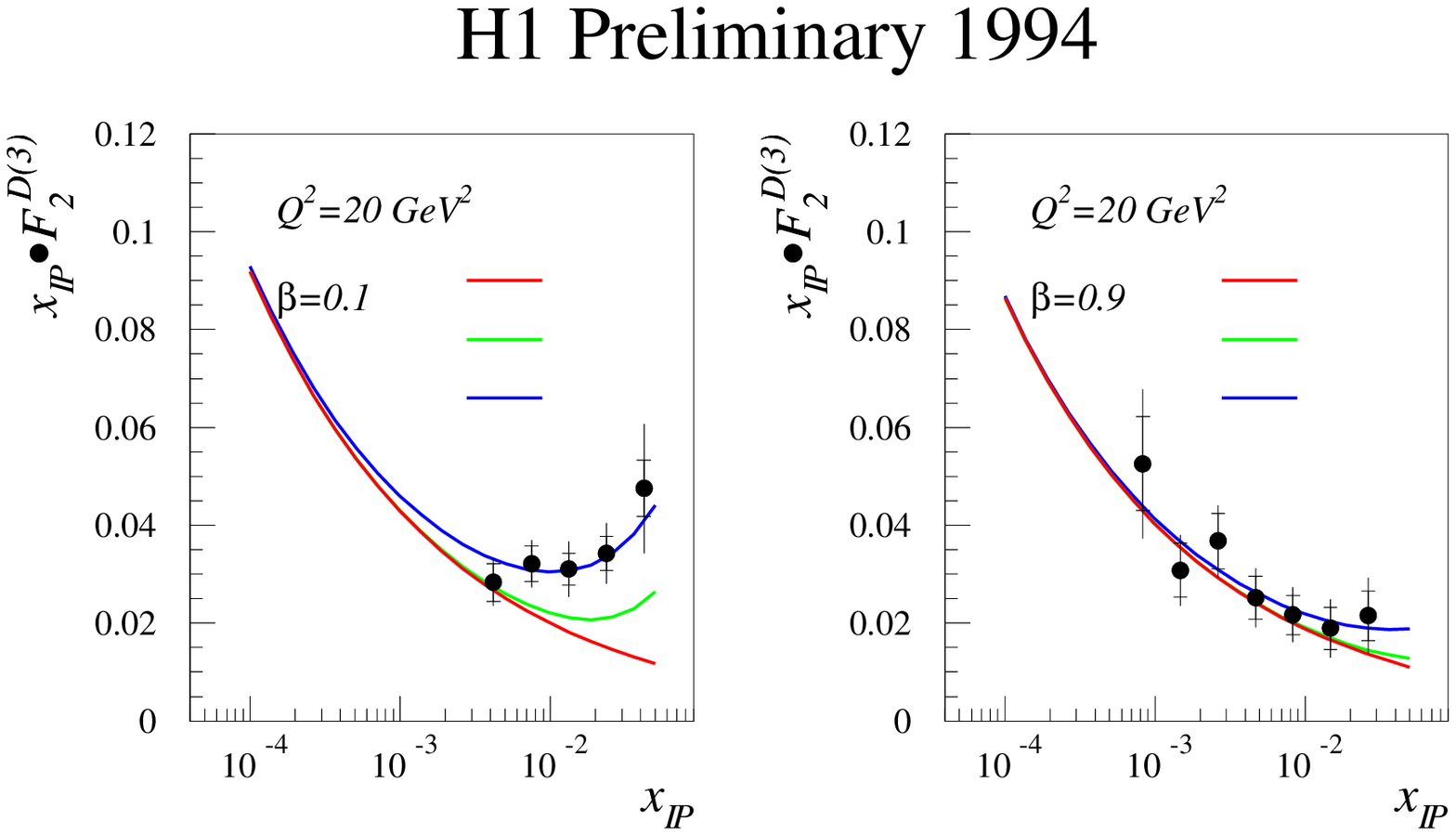,width=0.7\textwidth}}
     \put(100,2){\epsfig{file=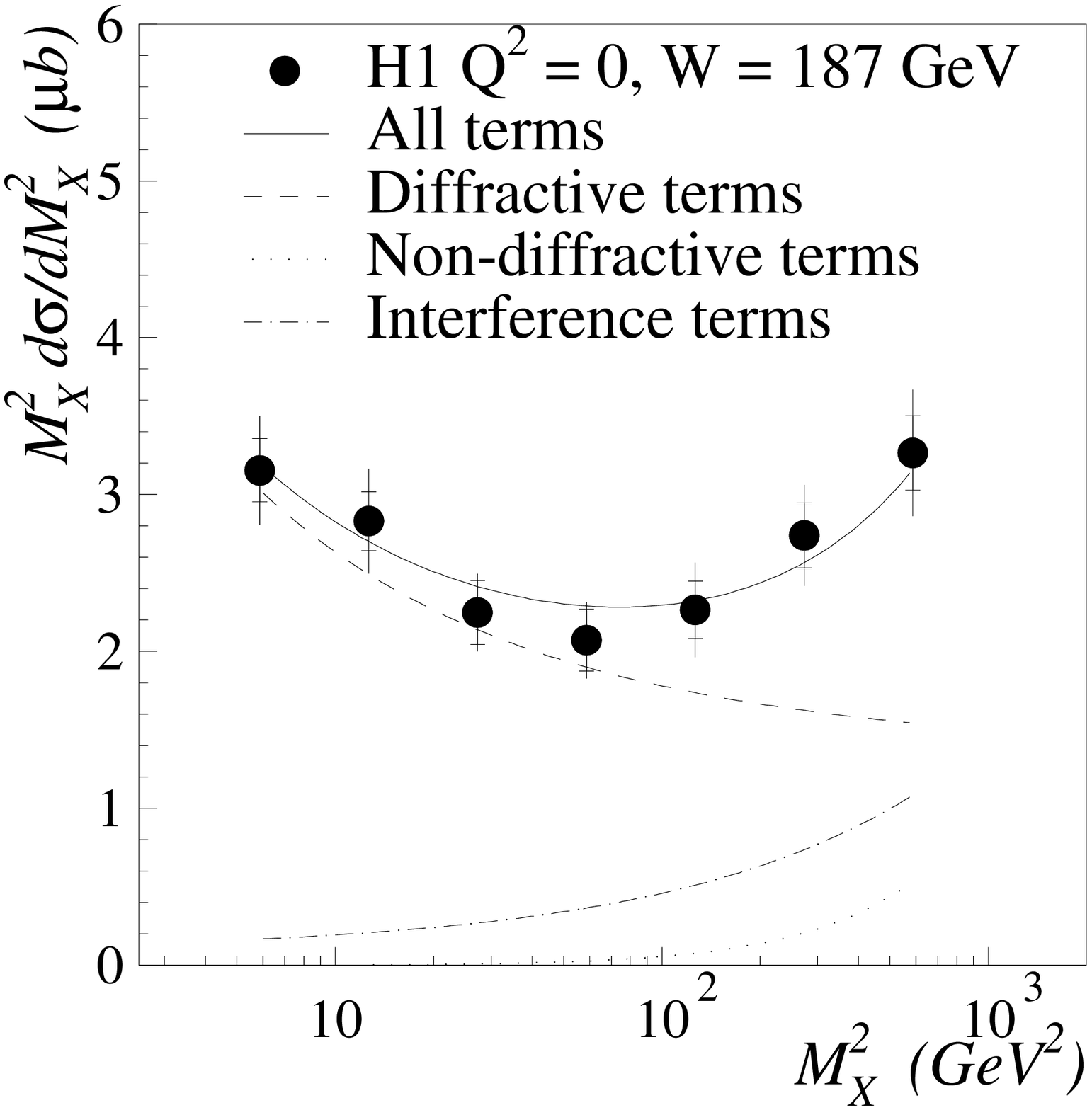,width=0.3\textwidth}}
     \put(17,12){\bf (a)}
     \put(60,12){\bf (b)}
     \put(107,12){\bf (c)}
   \end{picture}
 \end{center}
 \scaption {(a) and (b) Examples $\xpom$ dependences of $\xpom \ftwod$ at
two fixed values of $\beta$ and $Q^2$, such that the $\xpom$ variation is
generated by a variation in $W$. The results of the fit to 
equation~\ref{prn:f2dfac} with two interfering trajectories are superimposed.
The three lines shown represent the diffractive, diffractive $+$ interference
and the sum of all contributions. (c) The $\mx^2$ dependence of the 
photoproduction cross section 
$\mx^2 {\rm d} \sigma / {\rm d} \mx^2$ at fixed $W$,
together with the results of a fit with a similar decomposition of the cross
section in a triple Regge model.}
 \label{prn:regge}
\end{figure}

In a fit to the full data 
sample\cite{prn:dirkmann}, H1 find that a description 
of $F_2^{D(3)}$ with a diffractive exchange alone
is only viable if the trajectory intercept $\alphapom(0)$ has a $\beta$ 
dependence. There is no evidence of any need for a $Q^2$ dependence in the
present kinematic range of the measurement. This breaking of factorisation in
the measured cross section may be explained naturally
without the need for a $\beta$
dependent intercept by introducing a second trajectory with lower intercept.
The resulting fit to the data is good, with 
$\chi^2 / {\rm n.d.f.} = 165 / 156$ if the two contributing trajectories
interfere and $\chi^2 / {\rm n.d.f.} = 170 / 156$ if they don't. In
the kinematic range studied, the data are well described in the two reggeon
model with a leading singularity of intercept 
$\alphapom(0) = 1.18 \ \pm \ 0.02 \ ({\rm stat.}) \ \pm 0.04 \ ({\rm syst.})$,
a little larger than that describing hadron-hadron 
interactions\cite{prn:dlstot}. The secondary trajectory has intercept
$\alphareg(0) = 0.6 \ \pm \ 0.1 \ ({\rm stat.}) \  \pm 0.3 \ ({\rm syst.})$,
consistent with the approximately degenerate $\rho$, $\omega$ $f$ and $a$
trajectories that are also required to describe total cross 
sections\cite{prn:dlstot}. 
Since the pomeron structure function has rather a flat $\beta$ 
dependence\cite{prn:h1f2d93,prn:zeus1} and trajectories related to mesons
with lower intercepts than the pomeron have structure functions falling 
rapidly with $\beta$\cite{prn:grvpi}, equation (\ref{prn:f2dfac}) 
implies that the sub-leading
contribution should be most important at small $\beta$ and large $\xpom$. 
This can clearly be seen from figure~\ref{prn:regge}a and b, where
the $\xpom$ (or 
equivalently $W$) dependence is shown at two fixed values of $\beta$ and $Q^2$.

ZEUS use three different methods for the extraction of diffractive cross
sections. The 1993 $F_2^{D(3)}$ data\cite{prn:zeus1} discussed in 
section~\ref{prn:jets} were extracted using a large rapidity gap selection.
The second method uses the fact that
the ZEUS leading proton detectors are sensitive in the low $\xpom$
region, leading to a direct measurement of the single dissociation process
$Y = p$ and allowing a measurement of the differential $t$
distribution. In the kinematic range, $5 < Q^2 < 20 \ {\rm GeV^2}$,
$0.03 < y < 0.8$, $0.015 < \beta < 0.5$ and $\xpom > 0.03$ the data are
consistent with an exponential dependence $e^{b t}$, with slope parameter
$b = 7.1 \ \pm 1.1 \ ({\rm stat.}) \ ^{+0.7}_{-1.0} \ ({\rm syst.}) \ 
{\rm GeV^{-2}}$\cite{prn:grothe}. It will be very interesting in the future 
to see whether this figure shows any dependence 
on $\xpom$ in order to determine whether 
there is any shrinkage of the forward peak in diffractive DIS and to extract
the relevant value of $\alphapom^{\prime}$.

\begin{figure}[h] \unitlength 1mm
 \begin{center}
   \begin{picture}(120,37.5)
     \put(0,-3.75){\epsfig{file=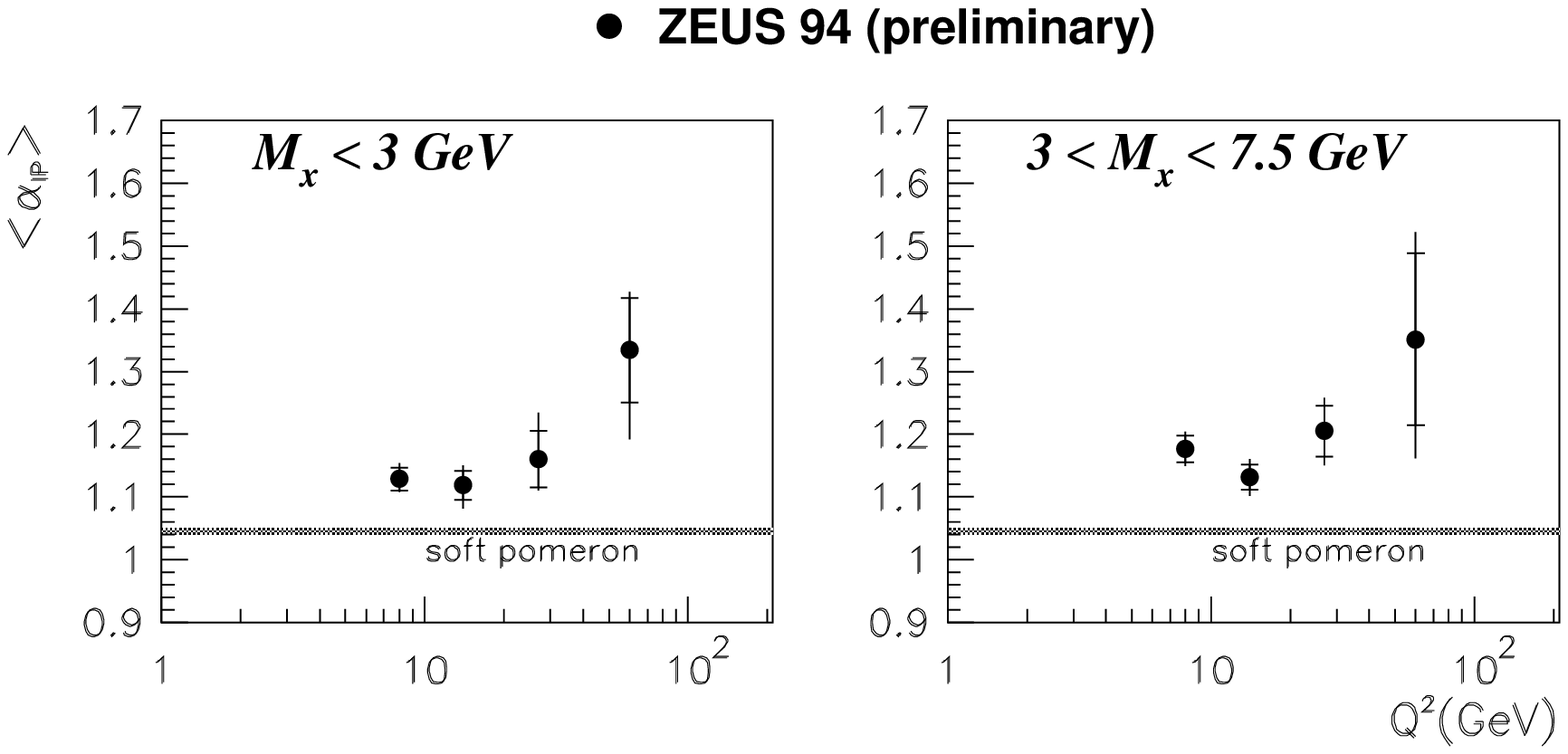,width=0.75\textwidth}}
     \put(13.8525,10.275){\dashbox{1.0}(35.8125,3.75)}
     \put(59.9025,10.275){\dashbox{1.0}(35.8125,3.75)}
   \end{picture}
 \end{center}
 \vspace{0.3cm}
 \scaption {The $Q^2$ dependence of the value of the pomeron trajectory
averaged over the $t$ distribution of the cross section in the ZEUS mass
subtraction analysis. A $1 \sigma$ error band (dashed
lines) for the prediction from the H1 fits to 
equation~\ref{prn:f2dfac} is superimposed.}
 \label{prn:zeusalphapom}
\end{figure}

The third method employed by ZEUS\cite{prn:briskin} is based on the fact that
different exchanges give rise to different $\mx$ distributions at fixed $W$
and $Q^2$. In bins of $W$ and $Q^2$, the raw measured $\mx$ distribution
with $\my < 4 \ {\rm GeV}$ is subjected to a fit of the form 
$\mx^2 {\rm d} \sigma / {\rm d} \mx^2 = D + N (\mx^2)^b$, with the 
normalisations $D$ and $N$ and the slope $b$ as free parameters. The constant
term $D$ is operationally defined as the diffractive contribution. This 
is similar to a triple Regge model (see below) 
with a triple pomeron
contribution and a single effective non-diffractive term\footnote{In general 
Regge theory predicts more complicated non-diffractive 
and diffractive contributions than are
allowed by the two terms in this parameterisation.}.
The pomeron intercept is extracted from the $W$ dependence of the 
diffractive contribution $D$ at fixed $Q^2$. The results, shown in 
figure~\ref{prn:zeusalphapom}, give an indication of a $Q^2$ dependence of the
pomeron trajectory. 
However, the uncertainties are large and when the ZEUS data are
compared with the 
$1 \sigma$ error band on the H1 result (dashed lines), the two experiments
are found to be in reasonable agreement. Comparisons on a point by point
basis between the ZEUS results obtained in the mass subtraction and leading 
proton measurements\cite{prn:grothe} do not reveal any large differences
beyond those expected from the proton dissociative contribution.
The leading proton measurement leads to a value of the diffractive
trajectory averaged over the $t$ distribution of the cross section 
of $\langle \alphapom(t) \rangle = 1.02 \ \pm 0.05 \ 
({\rm stat.}) \ \pm 0.07 ({\rm syst.})$, assuming a diffractive contribution 
only. This lies somewhat lower than the results shown in 
figure~\ref{prn:zeusalphapom}. The
difference in the measured values of $\langle \alphapom \rangle$ is likely 
to be
a consequence of the substantial sub-leading contributions that H1 find
to be needed in the kinematic regime of the leading proton measurement.

In the photoproduction regime, both H1\cite{prn:mxpaper} and
ZEUS\cite{prn:briskin} present results in the form of the
cross section 
${\rm d} \sigma^{\gamma p \rightarrow X Y} / {\rm d} \mx^2$ at fixed values of 
$W \sim 200 \ {\rm GeV}$, such that $\mx^2$ is approximately proportional to
$\xpom$. The H1 results are integrated over 
$|t| < 1.0 \ {\rm GeV^2}$ and $\my < 1.6 \ {\rm GeV}$. The ZEUS results
are integrated over $\my < 2.0 \ {\rm GeV}$ and all $t$.
A triple Regge approach is taken by both collaborations, such that both the
$\mx$ and the $W$ dependence of the cross section can be parameterised 
as\cite{prn:mxpaper,prn:reviews}
\begin{eqnarray}
  \frac{ {\rm d} \sigma}{{\rm d}t \, {\rm d}\mx^2} = \frac{s_{_0}}{W^4} 
  \sum_{i,j,k} G_{ijk}(t) \
  \left(\frac{W^2}{\mx^2}\right)^{\alpha_i(t) + \alpha_j(t)} 
  \left(\frac{\mx^2}{s_{_0}} \right)^{\alpha_k(0)} \hspace*{-0.1cm}
  \cos \left[ \phi_i(t) - \phi_j(t) \right] \ ,
\end{eqnarray}
where $i$ and $j$ correspond to the physical reggeon coupling to the proton
($i \neq j$ only for interference terms) and $k$ is a further reggeon 
describing the forward elastic amplitude 
$\gamma \alpha_i \rightarrow \gamma \alpha_j$ at a centre
of mass energy given by $\mx$. The hadronic mass scale $s_{_0}$ is
customarily taken to
be $1 \ {\rm GeV^2}$, $\phi_i(t)$ is the phase of reggeon $i$, completely
specified by the signature factor, and $G_{ijk}(t)$ contains all of the
couplings in the triple Regge amplitude $ijk$. 

H1 include fixed target data\cite{prn:E612} in fits that decompose the hadron 
level cross section into diffractive and non-diffractive triple Regge 
amplitudes. The differential photoproduction cross section from H1
is shown in figure~\ref{prn:regge}c. As can be seen from the shape of the 
$\mx^2$ spectrum, non-diffractive contributions become significant at large
$\mx^2$.
The increased importance of sub-leading contributions with
decreasing $\beta$ in the DIS regime, as explained above
from structure function
arguments, also implies that the sub-leading terms should increase in 
importance as $\mx$ increases. In photoproduction, the same effect can be
understood from Regge arguments alone. ZEUS fit a 
triple pomeron term only to the data with $8 < \mx < 24 \ {\rm GeV}$
after a Monte Carlo subtraction of non-diffractive contributions.
There is agreement between the two experiments that 
the pomeron intercept extracted
from these data is compatible with that describing hadron-hadron and
photoproduction elastic and total cross sections at high energy. H1 obtain
$\alphapom(0) = 1.068 \pm 0.016 \ {\rm (stat.)} \pm 0.022 \ {\rm
(syst.)} \pm 0.041 \ {\rm (model)}$, where the model dependence error is
dominated by uncertainties in the details of the sub-leading terms.
The ZEUS result is
$\alphapom(0) = 1.12 \pm 0.04 \ {\rm (stat.)} \pm 0.08 \ {\rm (syst.)}$.

ZEUS use their leading proton spectrometer to measure the differential $t$
distribution for photoproduction
in the kinematic range $0.07 < |t| < 0.4 \ {\rm GeV^2}$,
and $4 < \mx < 32 \ {\rm GeV}$ at $\av{W} \sim 200 \ {\rm GeV}$. The results
are well described by an exponential parameterisation with slope parameter
$b = 7.3 \pm 0.9 \ {\rm (stat.)} \pm 1.0 \ {\rm (syst.)} \ {\rm GeV^{-2}}$, a
little larger than that measured at 
$\av{W} \sim 14 \ {\rm GeV}$\cite{prn:E612} and consistent with shrinkage
of the forward diffractive peak with 
$\alphapom^{\prime} \sim 0.25 \ {\rm GeV^{-2}}$.
Studies by both collaborations\cite{prn:mxpaper,prn:zeusmx,prn:briskin} of
the diffractive contribution to the total photoproduction cross section
reveal that unitarity bounds\cite{prn:pumplin} 
are approached to within a factor of 2.

\section{The Deep Inelastic Structure of Colour Singlet Exchange at low 
{\boldmath $\xpom$}}
\label{prn:pdfs}

H1 integrate $F_2^{D(3)}$ over a fixed range in $\xpom$ to form the quantity
\begin{eqnarray}
  \tilde{F}_2^D (\beta, Q^2) = \int_{\xpom^{\rm low}}^{\xpom^{\rm high}}
\ftwod \ {\rm d} \xpom
\label{f2dtilde}
\end{eqnarray}
with limits $\xpom^{\rm low} = 0.0003$ and $\xpom^{\rm high} = 0.05.$ In
the factorisable prescription of Ingelman and Schlein\cite{prn:is},
$\tilde{F}_2^D$ provides a measurement of the deep inelastic structure of the
pomeron. In the scenario in which more than one exchange contributes to the
cross section, $\tilde{F}_2^D$ is an effective structure function for whatever
exchanges contribute in the $\xpom$ and $t$ ranges of the measurement.
Scaling violations
with positive $\partial \tilde{F}_2^D / \partial \log Q^2$ are found to persist
to values of $\beta$ of at least $0.65$\cite{prn:dirkmann}, indicating the
need for substantial gluon structure with momentum fractions $x_{g/\pom}$
greater than this figure at low $Q^2$. These conclusions are not affected when
the phenomenological fits described in section~\ref{prn:incl} are used to
subtract the non-diffractive contributions from $\tilde{F}_2^D$.

\begin{figure}[h]
   \vspace*{-0.1cm}
   \begin{center}
   \epsfig{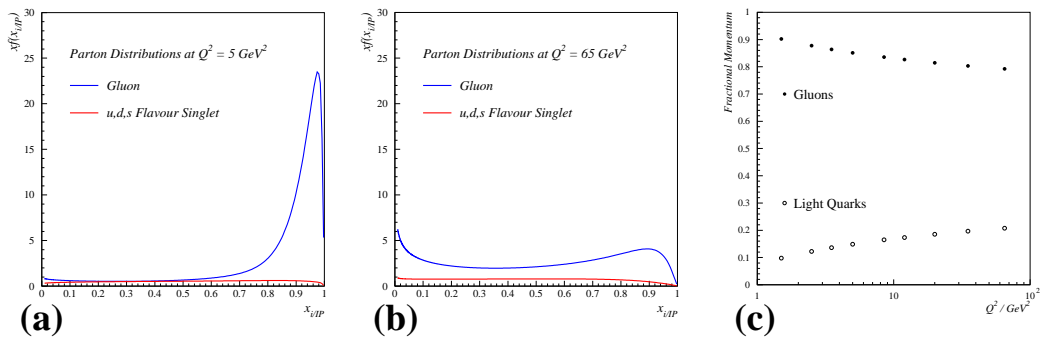}
   \end{center}
   \vspace*{-0.2cm}
    \caption{Best fit momentum weighted quark and gluon distributions in 
fractional momenta $x_{g/\pom}$ and  $x_{q/\pom}$ for the
exchange averaged over $\xpom$ and $t$ at
(a) $Q^2 = 5\ {\rm GeV^2}$ and (b) $Q^2 = 65\ {\rm GeV^2}$;
c) fraction of the total momentum transfer carried by quarks and by gluons as
a function of $Q^2$.}
\label{prn:pdf}
\vspace{-0.2cm}
\end{figure}

Subjecting the $\tilde{F}_2^D$ 
data to a QCD analysis in which the pomeron parton distributions
evolve according to the leading order DGLAP evolution 
equations\cite{prn:dglap}, the best solution is a gluon 
distribution strongly peaked as
$x_{g/\pom} \rightarrow 1$ at low $Q^2$ 
(figure~\ref{prn:pdf}a and b)\footnote{There was much discussion at the 
workshop on the question of whether the DGLAP approach is appropriate for 
$\tilde{F}_2^D$ at all $\beta$ and $Q^2$. A 
summary can be found in\cite{prn:soper}.}. Just how
peaked the gluon distribution has to be is not yet fully determined.
The conclusion holds true
when the data at $\beta = 0.9$, corresponding to the vector meson resonance
region, are omitted from the fit. In
excess of $80 \ \%$ of the pomeron momentum is found to be carried by gluons
throughout the $Q^2$ range studied (figure~\ref{prn:pdf}c). No good solutions
have been obtained in which the exchange structure is dominated by quarks.

\section{The Final State System {\boldmath $X$} at low {\boldmath $\xpom$}}
\label{prn:fs}

\begin{figure}[h] \unitlength 1mm
 \begin{center}
   \begin{picture}(160,35)
     \put(0,3){\epsfig{file=prn.qpm.epsf,width=0.25\textwidth}}
     \put(56,3){\epsfig{file=prn.bgf.epsf,width=0.22\textwidth}}
     \put(106,3){\epsfig{file=prn.qcdc.epsf,width=0.25\textwidth}}
     \put(0,16){\Large{\bf (a)}}
     \put(53,16){\Large{\bf (b)}}
     \put(106,16){\Large{\bf (c)}}
   \end{picture}
 \end{center}
 \vspace{-0.3cm}
 \scaption {Example zeroth and first order QCD processes contributing to 
diffractive DIS in the model in which a distinct set of diffractive parton
distributions is probed. (a) ${\cal O} (\alpha_{\rm em})$ `quark parton model'
diagram; (b) ${\cal O} (\alpha_{\rm em} \alpha_s)$ `boson-gluon fusion'
diagram; (c) ${\cal O} (\alpha_{\rm em} \alpha_s)$ `QCD-Compton' diagram. 
Up to leading order, only
the boson-gluon fusion process is initiated by a gluon from the parton
distributions.}
 \label{prn:partons}
\end{figure}
In the factorisation model (see section~\ref{prn:incl}) in which
diffractive DIS is viewed as deep-inelastic scattering from distinct 
diffractive
parton distributions, 
the lowest order QCD process by which a quark couples to
a photon is the ${\cal O} (\alpha_{\rm em})$ `quark-parton
model' diagram (figure~\ref{prn:partons}a). The lowest order process by 
which a gluon can couple to the photon is the 
${\cal O} (\alpha_{\rm em} \alpha_s)$ boson-gluon fusion process
(figure~\ref{prn:partons}b). These are expected to be the 
dominant parton
level interactions in the quark and gluon dominated pomeron scenarios
respectively. These two diagrams lead to rather different observable
characteristics of the system $X$. The
natural frame in which to study the final state is the rest frame of the system
$X$ (or equivalently the centre of mass of the $\gamma^{\star} \pom$
interaction). The natural axis in this frame is that of the interacting
photon and pomeron. Several final state 
analyses were presented by H1 and ZEUS
which corroborate the conclusion from the QCD analysis of the inclusive
cross section (section~\ref{prn:pdfs})
that the pomeron in 
hard diffractive scattering must contain a large fraction of gluons with 
large $x_{g/\pom}$. 

\begin{figure}[h] \unitlength 1mm
 \begin{center}
   \begin{picture}(160,62)
     \put(0,0){\epsfig{figure=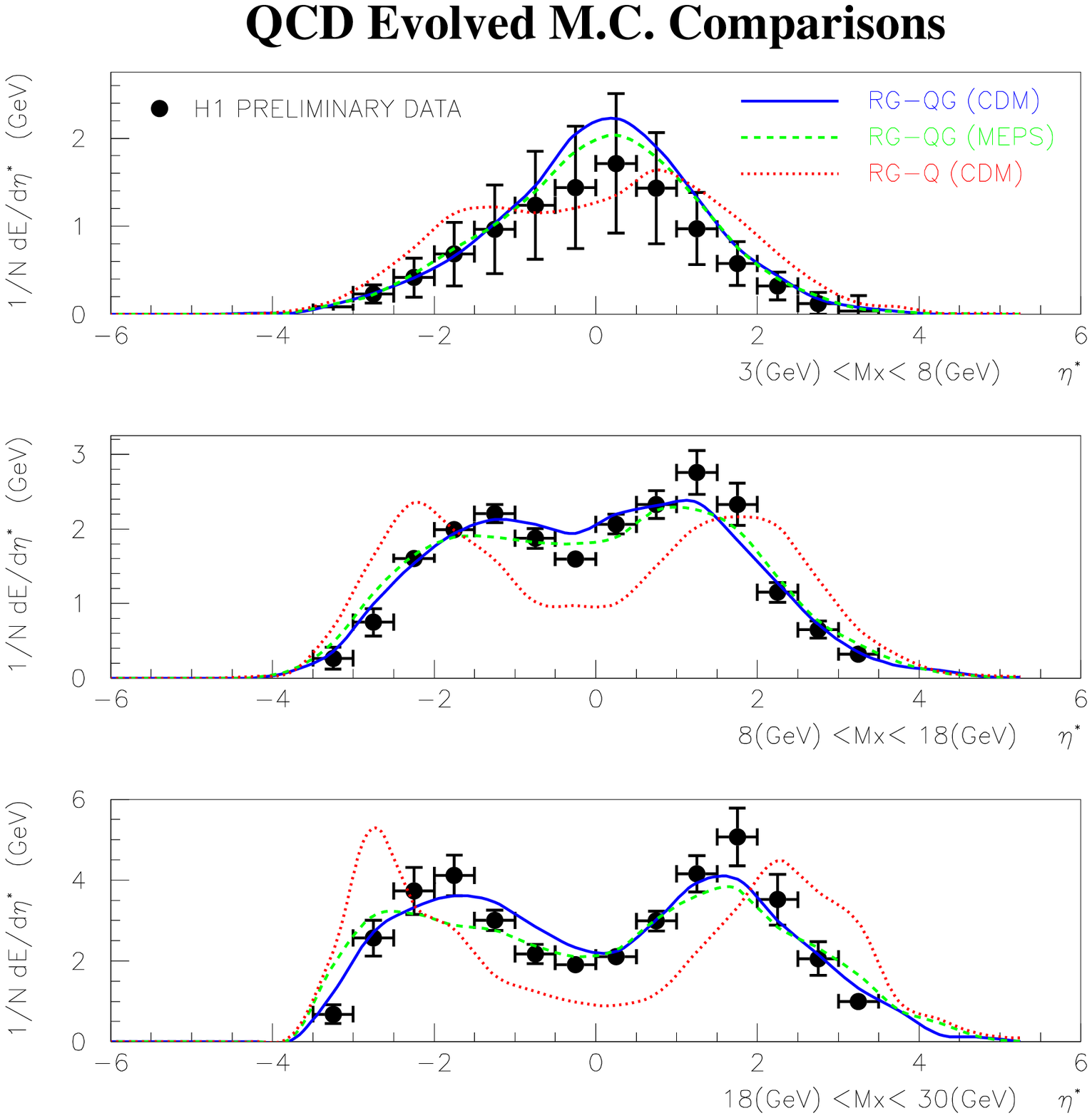,width=0.5\textwidth}}
     \put(72,0){\epsfig{figure=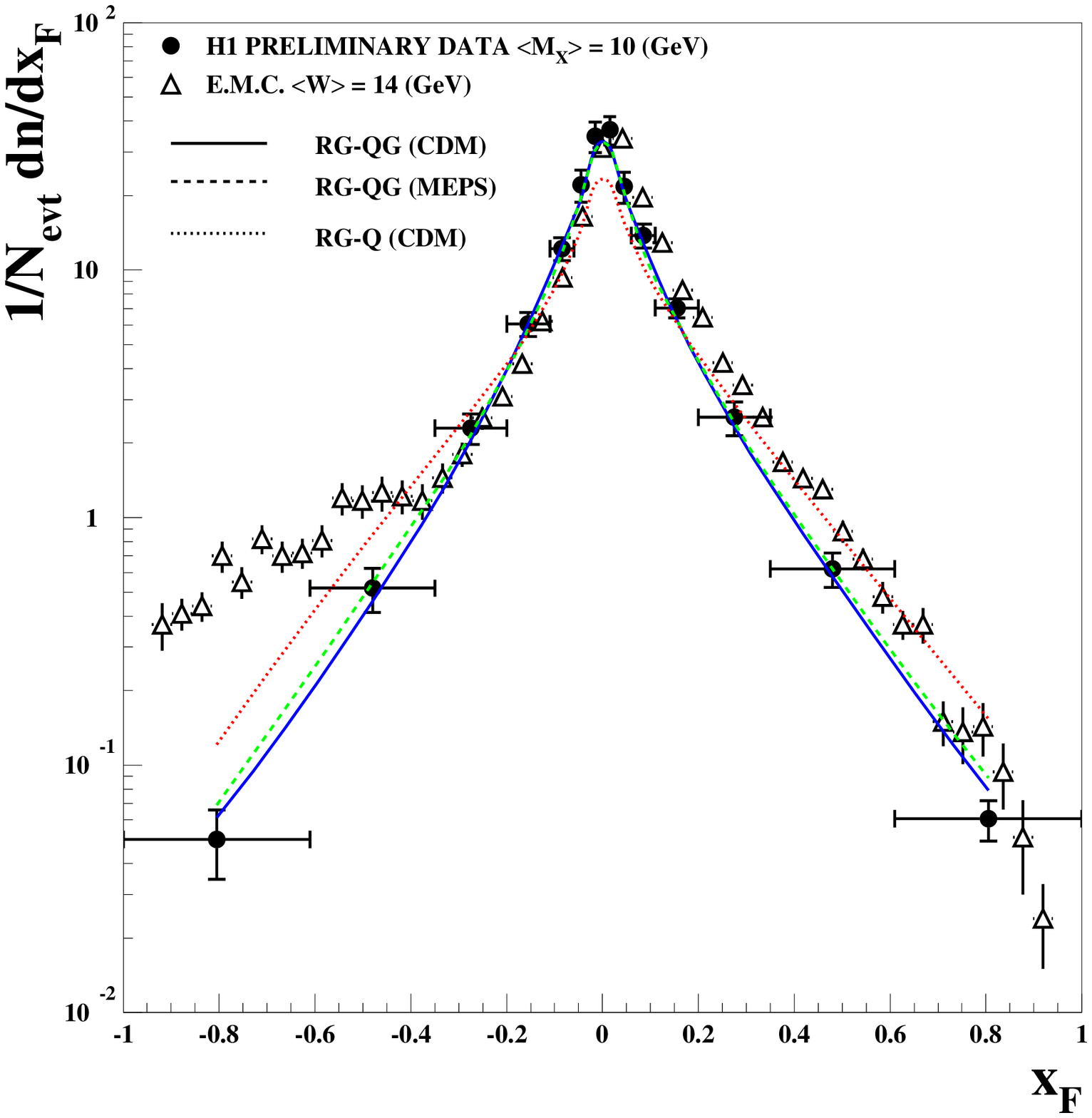,width=0.5\textwidth}}
     \put(0,-3){\Large{\bf (a)}}
     \put(77,-3){\Large{\bf (b)}}
     \put(33,-3){\vector(-1,0){10}}
     \put(37,-3){\vector(1,0){10}}
     \put(106,-3){\vector(-1,0){10}}
     \put(110,-3){\vector(1,0){10}}
     \put(18,-3.7){\boldmath $\pom$}
     \put(48,-3){\boldmath $\gamma^{\star}$}
     \put(91,-3.7){\boldmath $\pom$}
     \put(121,-3){\boldmath $\gamma^{\star}$}
   \end{picture}
 \end{center}
 \vspace{0.4cm}
 \scaption {(a) Energy flow in the rest frame of the system $X$ in
intervals of $\mx$, compared to two Monte Carlo models (RG-QG) containing the 
implementation of parton distributions from the best QCD fits obtained to
$\tilde{F}_2^{D} (\beta, Q^2)$ and to a Monte Carlo model (RG-Q) with a quark
dominated pomeron. (b) $x_{_{\rm F}}$ distribution in the same frame compared
to the same models and to inclusive proton DIS 
data\protect \cite{prn:emc} \protect at a value
of $W$ similar to $\mx$ in the diffractive data. In each plot the distributions
are made relative to the $\gamma^{\star} \pom$ collision axis.}
 \label{prn:eflow}
\end{figure}

If the 
pomeron consists dominantly of quarks and the total diffractive cross section
is dominated by the QPM process, large momenta transverse to the
$\gamma^{\star} \pom$ axis are expected to arise at highest order 
from the ${\cal O} (\alpha_s)$ suppressed QCD-Compton process (e.g. 
figure~\ref{prn:partons}c). Other sources, such as
intrinsic transverse momentum of partons in the pomeron are expected to be
small. By contrast,
in the boson-gluon fusion process, the quark propagator can be highly
virtual, giving rise to substantial high $\pt$ particle production
and energy flow in the central
rapidity region of the $\gamma^{\star} \pom$ centre of mass frame.
H1 have measured energy flow, charged particle transverse momentum spectra,
$x_{_{\rm F}}$ spectra\footnote{The Feynman variable is defined here as 
$x_{_{\rm F}} = 2 p_z^{\star} / \mx$, where $p_z^{\star}$ is the longitudinal
momentum of each particle relative to the $\gamma^{\star} \pom$ axis.} and the
mean $\pt$ as a function of $x_{_{\rm F}}$ (`seagull plot') in this
frame\cite{prn:cormack}. The energy flow distributions are shown in three
bins of $\mx$ in figure~\ref{prn:eflow}a and 
are compared with Monte Carlo
simulations\cite{prn:rapgap} that 
incorporate two sets of evolving
parton distributions for the pomeron. The first,
labelled RG-QG in the figures, corresponds to 
the best QCD fits to $F_2^{D(3)}$ in which
the pomeron parton distributions are as shown in figure~\ref{prn:pdf}. A 
sub-leading exchange is also included as obtained from the fits to
$F_2^{D(3)}$. Two different fragmentation schemes, `MEPS' and
`CDM' are used. To demonstrate the sensitivity of final state
observables to the parton distributions, a second set, corresponding to the
best fit to the $F_2^{D(3)}$ data in which the pomeron
consists only of quarks at low $Q^2$,
is also implemented (labelled RG-Q in the figures). There is considerable 
sensitivity to the difference between the RG-QG and RG-Q simulations,
with a good description being obtained with the
leading gluon parton distributions in either fragmentation scheme. 
The quark dominated model RG-Q does not predict enough
energy flow in the central region. 

The $x_{_{\rm F}}$ spectra for charged particles (figure~\ref{prn:pdf}b)
are compared with the three Monte Carlo simulations and
also with inclusive DIS data\cite{prn:emc}
at a mean value of $W$ similar to the mean
$\mx$ in the diffractive data. The pomeron and photon hemispheres 
are found to be rather symmetric in the $x_{_{\rm F}}$ distribution, as can 
also be see from the energy flow distributions. This behaviour is well 
described by the RG-QG models and is to be expected in
the scenario where the pomeron parton distributions are dominated by gluons
carrying large $x_{g/\pom}$. The interaction is then essentially 
$\gamma^{\star} g \rightarrow q \bar{q}$, with the pomeron `remnant' carrying
only a small momentum fraction. The symmetry between the two hemispheres in
the diffractive data is in marked contrast with the inclusive DIS data,
where there is an extended proton remnant, leading to an enhancement in the
$x_{_{\rm F}}$ distribution in the proton fragmentation region. If DGLAP is
appropriate for describing the evolution of diffractive parton distributions,
a remnant would be expected to become increasingly visible with increasing
$Q^2$ (compare figures~\ref{prn:pdf}a and b). It will be interesting to test 
this hypothesis with future data.

Similar conclusions are obtained in analyses of event shapes. H1 study 
thrust\cite{prn:cormack} and ZEUS sphericity\cite{prn:hernandez}. These are
particularly important analyses, since the low masses accessed in
diffraction at HERA severely limit the phase space for jet production, but
jet-like structures can still be resolved from event shapes.
Both experiments find that the final state system $X$ becomes increasingly
collimated along the $\gamma^{\star} \pom$ axis as $\mx$ increases, 
consistent with a dominant two jet-like configuration with hadronisation 
effects broadening the jet structures. However, the
collimation of the diffractive
events is less pronounced than is the case in $e^+ e^-$ annihilation at 
centre of mass energy equal to $\mx$, demonstrating that the system
$X$ is more complex than a simple $q \bar{q}$ system modified
by QCD gluon radiation. Both the distribution in momentum of the 
thrust jets transverse to the $\gamma^{\star} \pom$ axis and the mean particle
momentum transverse to this axis show a significant tail to large values, 
indicating the need for substantial contributions from final states containing
more than two partons. The results from both experiments are well described by 
Monte Carlo models that are based on a hard gluon dominated pomeron structure.
Models with a quark dominated structure over-estimate the collimation
of the events and underestimate the fraction of events with high $\pt$
particles or thrust jets.

Correlations between the charged particle 
multiplicities in the photon and pomeron hemispheres and multiplicity moments
have been studied by H1\cite{prn:mechelen}. It is clear that the long
range correlations between hadrons in the two hemispheres is 
greater in the diffractive data than is the
case in $e^+ e^-$ data at similar centre of mass energy, demonstrating that
the colour connections between partons are more complex in the diffractive
case. Again, this indicates\cite{prn:strings} that the lowest order process
does not result in a final state quite as simple as $q \bar{q}$.
The observation is consistent with a final state consisting of a $q \bar{q}$
pair with an additional coloured pomeron `remnant' 
as would be expected for the boson-gluon fusion process shown in 
figure~\ref{prn:partons}b.

Both collaborations have investigated the charm content of the system $X$ by
measuring inclusive cross sections for $D^{\star}$ production. 
Analysis of inclusive DIS data\cite{prn:h1dstar} 
supports the hypothesis that the charm contribution to the proton structure
function is dominated by boson-gluon fusion and shows that above threshold,
production rates are large. For a quark induced hard process, charm can
only be produced from any intrinsic charm content of the 
pomeron. Both ZEUS\cite{prn:terron} and H1\cite{prn:cormack}
measure $D^{\star}$ cross sections that are consistent 
with those predicted in gluon based Monte Carlo
models. They are inconsistent at the
$2 \sigma$ level or greater
with quark based models that do not include intrinsic charm
in the pomeron sea, though in some models\cite{prn:nnncharm}
the pomeron parton distributions contain significant charm.

\section{Dijet Production at low {\boldmath $\xpom$}}
\label{prn:jets}

Diffractive dijet photoproduction in the system $X$ with $\xpom < 0.05$
has been studied by ZEUS and H1
with H1 also investigating dijet electroproduction\cite{prn:marage,prn:terron}.
These analyses are
predominantly sensitive to the ${\cal O} (\alpha_{\rm em} \alpha_s)$
boson-gluon fusion and QCD-Compton processes
shown in figures~\ref{prn:partons}b and c. If the pomeron parton distributions
are quark dominated, then the bulk of the cross section is taken up by the
zeroth order process shown in figure~\ref{prn:partons}a, with the QCD-Compton
process suppressed by ${\cal O} (\alpha_s)$. If the pomeron structure
is dominated by gluons, the boson-gluon fusion process dominates the cross
section. Since the measured diffractive DIS cross section 
(section~\ref{prn:incl})
is an input parameter, the Monte Carlo models predict substantially more
dijet production for a gluon dominated than for a quark dominated exchange.

\begin{figure}[h] \unitlength 1mm
 \begin{center}
   \begin{picture}(120,75)
     \put(-7.5,0){\epsfig{figure=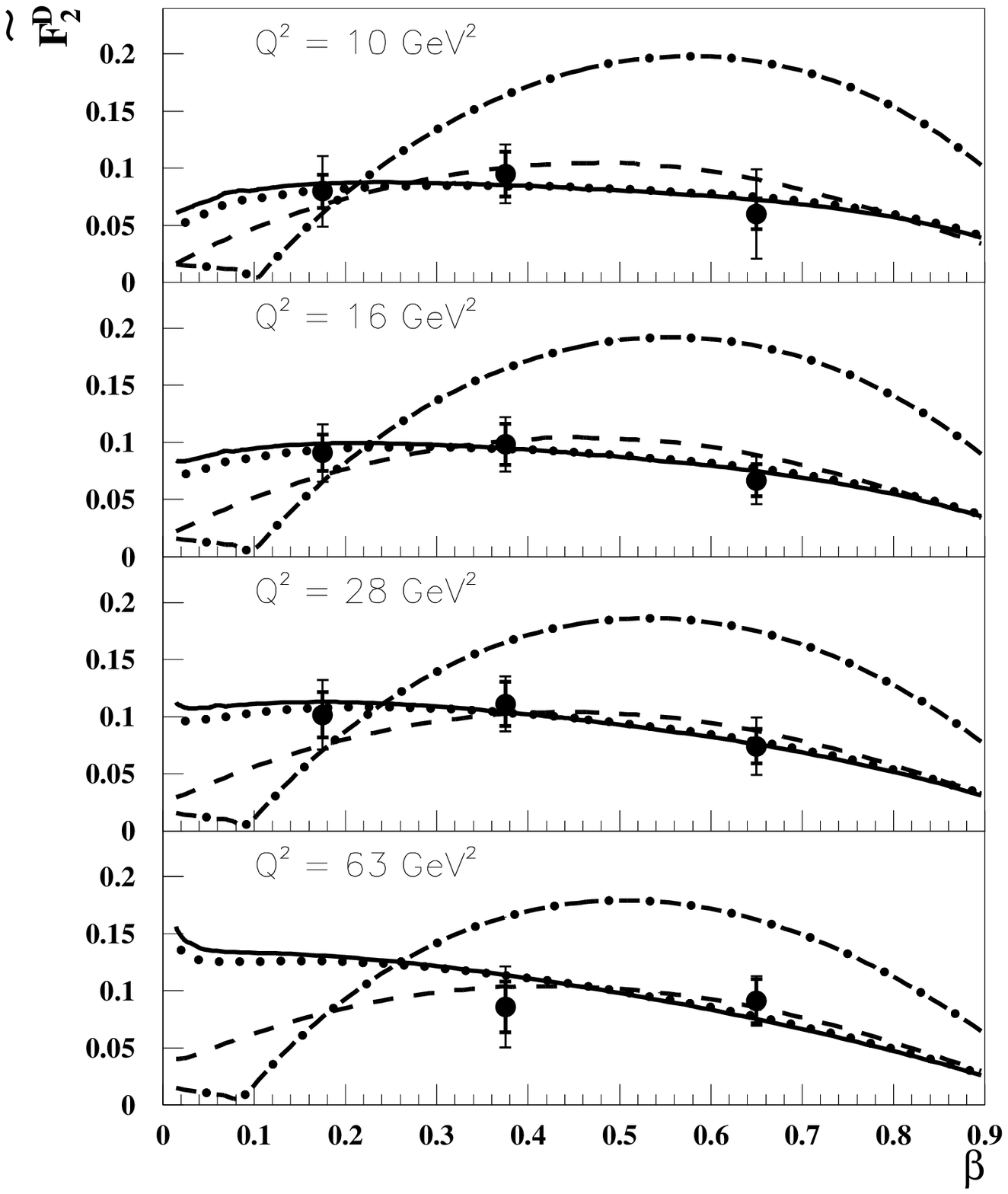,width=0.64\textwidth}}
     \put(64,41){\epsfig{figure=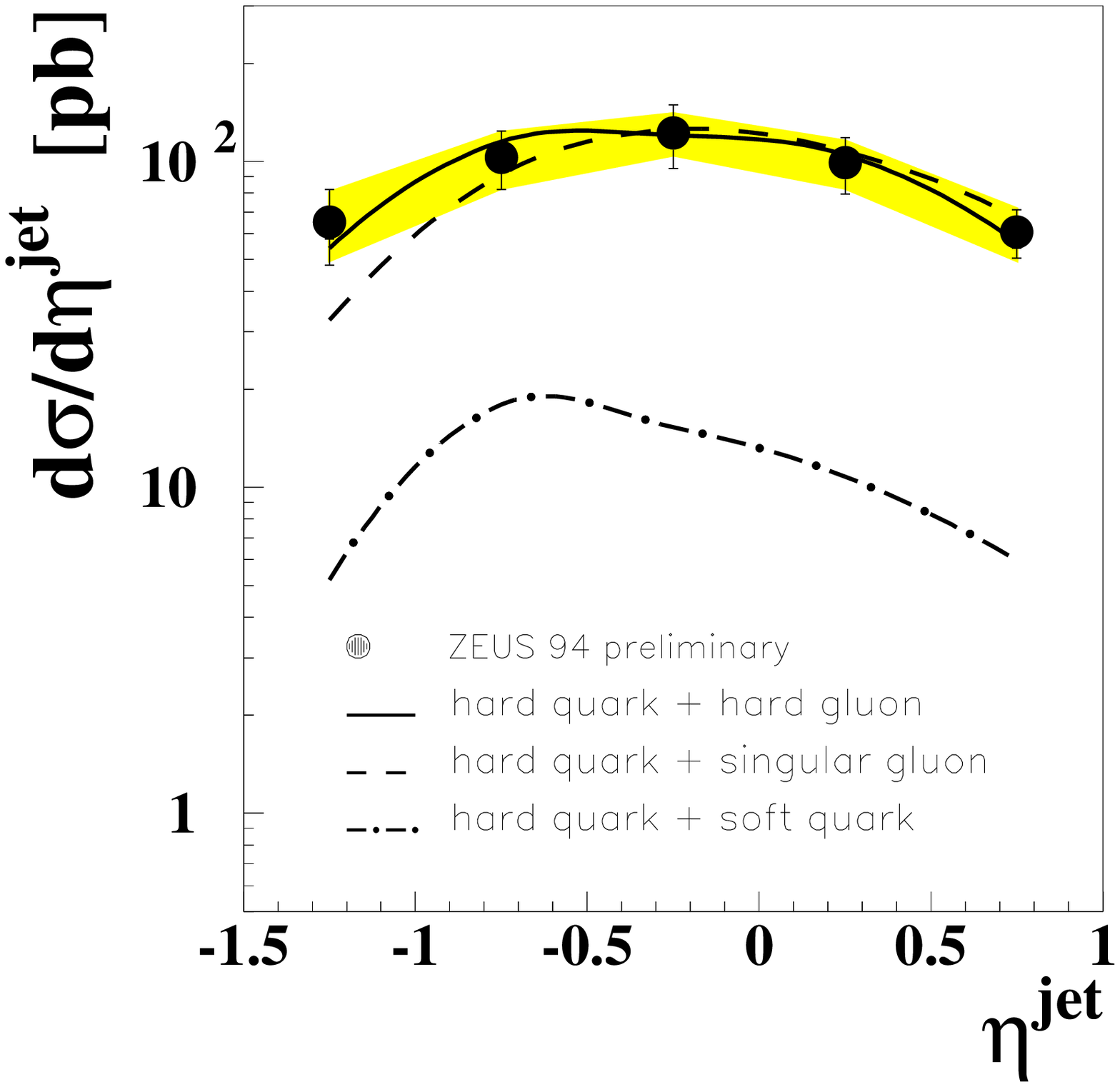,width=0.3\textwidth}}
     \put(67,0){\epsfig{figure=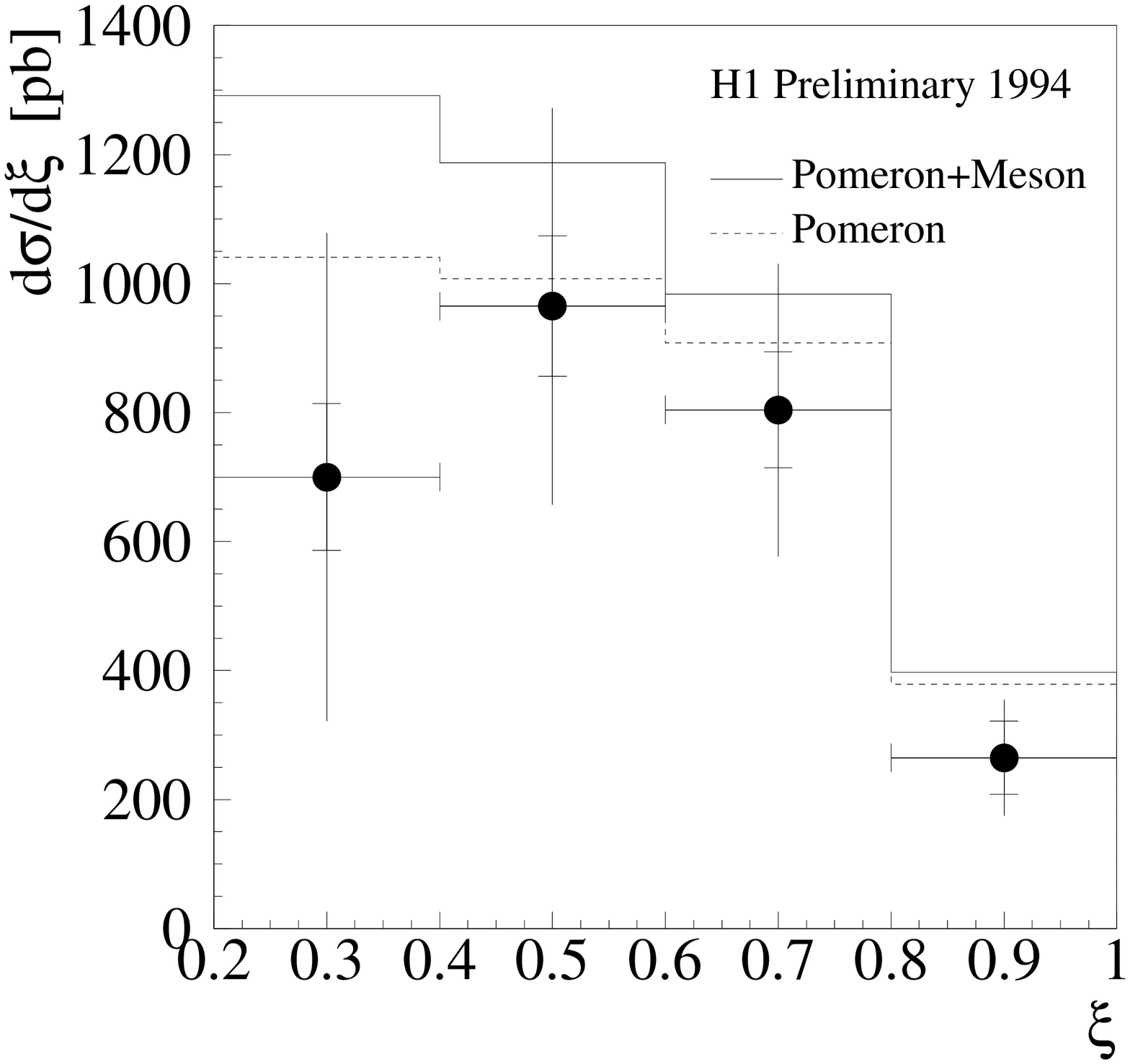,width=0.3\textwidth}}
     \put(-3,10){\bf (a)}
     \put(94,66){\bf (b)}
     \put(78,9){\bf (c)}
   \end{picture}
 \end{center}
 \vspace{-0.5cm}
 \scaption {Combined ZEUS fit to a) $\tilde{F}_2^D (\beta, Q^2)$ and b) the
pseudorapidity distribution of photoproduced dijets. The best DGLAP fits
with different parameterisations of pomeron parton densities, described in the
text, are also shown. The precise details of each parameterisation can be 
found in\protect \cite{prn:terron} \protect. (c) The H1 distribution in the
hadron level estimator $\xi$ of 
the momentum fraction of the exchange
entering the hard subprocess, compared to the RAPGAP model with pomeron
parton distributions derived from $F_2^{D(3)}$ and an additional 
sub-leading exchange.}
 \label{prn:zeusfit}
\end{figure}

ZEUS have performed a combined fit\cite{prn:terron} in 
the DGLAP framework to the quantity
$\tilde{F}_2^D$ (figure~\ref{prn:zeusfit}a)\cite{prn:zeus1} defined in 
equation~\ref{f2dtilde}, and the
pseudo-rapidity distribution of photoproduced dijets 
(figure~\ref{prn:zeusfit}b).
This approach assumes a universality of the
product of the pomeron flux and parton distributions and neglects any effects
due to imperfect rapidity gap survival probability\cite{prn:survive}. The
latter assumption may become invalid in the resolved photoproduction regime,
though any effect on the results of the fits is unlikely to be catastrophic 
since the data are dominated by the large $x_{\gamma}$ region\cite{prn:terron},
where the survival probabilities are expected to be largest.
The `hard quark $+$ hard gluon' and `hard quark $+$ singular gluon'
parameterisations both give good fits, yielding $87 \%$ and $69 \%$
gluon composition for the exchange respectively at the starting scale of
$4 \ {\rm GeV^2}$.  The sensitivity to how hard the gluon distribution
must be is limited.
The parameterisation with quarks only at the starting scale does not describe
$\tilde{F}_2^D$ and seriously underestimates the jet rates.

A similar procedure is followed in\cite{prn:whitmore}, where parton
distribution functions extracted from the $\tilde{F}_2^D$ measurement shown
in figure~\ref{prn:zeusfit}a are used to predict rates of dijet and $W$
production at the Tevatron under the assumption of universality of the
diffractive parton distributions. The 
predictions substantially over-estimate the rates measured by 
CDF\cite{prn:melese} and D0\cite{prn:D0}, indicating a breaking of 
diffractive factorisation consistent with a rather small rapidity gap survival
probability $\sim 0.1$ 
in diffractive $p \bar{p}$ interactions. Further comparisons between hard
diffraction data from HERA and the 
Tevatron\footnote{A summary of results from the Tevatron presented at the
workshop can be found in\cite{prn:soper}.} 
are very important in order to fully investigate these effects.

H1 find that the dijet rates and distributions in photoproduction and at 
high $Q^2$ are reasonably well described by the Monte Carlo implementations
of the diffractive parton distributions
derived from $F_2^{D(3)}$ at a scale set by the jet $\pt^2$. 
Quark dominated pomeron models substantially
under-estimate the dijet production rate\cite{prn:marage}.
It is clear from the 
photoproduction dijet distributions that there are both resolved and direct 
photon contributions\cite{prn:marage,prn:terron}. The hadron level variable
$\xi = \frac{\sum_{\rm jets} (E + p_z)}{\sum_X (E + p_z)}$
is an estimator of the fraction of the exchanged momentum that
is transferred to the dijet system. The distribution in this variable, as 
measured by H1 in photoproduction, is shown in figure~\ref{prn:zeusfit}c. It is
clear that the dominant process does not involve the full momentum of the
exchange entering the hard process, which would be expected in models based on
the exchange of two gluons in a colour singlet configuration without higher 
order corrections\cite{prn:glue2}. There is no evidence for a super-hard
contribution of the kind reported by UA8\cite{prn:ua8}, 
though the UA8 data were at
larger values of $|t|$ than have been accessed to date at HERA.

\section{Leading Baryons at Large {\boldmath $\xpom$}}
\label{prn:highxl}

The direct study of leading proton and neutron 
production in the proton fragmentation
region in DIS at relatively large 
$\xpom$\footnote{In the interests of internal consistency, this variable is
still referred to as $\xpom$ here, although this does not imply that pomeron
exchange is the dominant process! In \cite{prn:list} the variable is
referred to as $\xi$ and in \cite{prn:cartiglia} it is equivalent
to $1 - x_L$.} has developed considerably in the
past year. The region in $\xpom$ studied goes way beyond that where
diffraction is expected to be the dominant process. Two 
models are presently available that attempt to describe
the data. The exchange of charged or neutral Reggeised 
pions\cite{prn:reggepi} is implemented
in the POMPYT\cite{prn:pompyt} and RAPGAP\cite{prn:rapgap} Monte Carlo 
generators, with the leading order GRV parameterisation of the pion structure
function\cite{prn:grvpi}. The $\xpom$ dependence of $pp$ data in a similar
region is well described by such 
models\cite{prn:reviews}. An alternative approach,
implemented in the Monte Carlo generator LEPTO6.5\cite{prn:lepto}, is to 
attempt to predict all aspects of the proton fragmentation region in terms
of string fragmentation, with soft interactions changing colour configurations
but not parton momenta and hence yielding rapidity gaps\cite{prn:sci}.

For the leading proton analysis, H1 measure a structure function 
$F_2^{\rm LP(3)}(x, Q^2, \xpom)$\cite{prn:list}, defined in a similar
manner to $\ftwod$, in the region $6.5 \times 10^{-5} < x < 6 \times 10^{-3}$,
$2 < Q^2 < 50 \ {\rm GeV^2}$ and $0.1 < \xpom < 0.25$, 
integrated over transverse
momenta of the scattered proton $\pt < 200 \ {\rm MeV}$. The resulting 
structure function shows a weak dependence on $\xpom$, a logarithmic rise with
$Q^2$ at fixed $x$ and a slight fall with increasing $x$ at fixed $Q^2$.
The dependence on $x$ and $Q^2$ is compatible with that of the inclusive 
proton structure
function in the same range of $x$ and $Q^2$. The fraction of DIS
events with energetic
leading neutrons is also found to be independent of the
variables $x$, $Q^2$ and the observed charged track 
multiplicity\cite{prn:cartiglia,prn:jansen}, all of
which are associated with the
photon fragmentation region or hadronic
plateau.
Even when the raw neutron energy distribution is compared with that from a 
sample of events originating from proton beam interactions with residual gas
in the beam pipe, there are no significant differences in 
shape\cite{prn:cartiglia}. 
The fraction of DIS events with a leading neutron with $\xpom < 0.5$ and
$|t| < 0.5 \ {\rm GeV^2}$ is measured to be $9.1 \ ^{+3.6}_{-5.7} \ \%$ by
ZEUS\cite{prn:cartiglia} and $7.8 \ ^{+3.0}_{-2.0} \ \%$ by 
H1\cite{prn:jansen}. Similar fractions are observed in a similar range for
leading proton production\cite{prn:cartiglia}.  

All features of the H1 leading
proton measurement are well predicted by the RAPGAP implementation of pion
exchange, except that the overall normalisation in the model is
too small by a factor of around 2\cite{prn:list}. 
The soft colour exchange model predicts the
absolute normalisation well, but fails to reproduce the scaling violations.
Both the reggeised pion exchange
and the soft colour interaction models give a good description of the leading
neutron energy spectra with $\xpom \lapprox 0.5 \ {\rm GeV}$.

ZEUS have investigated the $t$ distribution of leading protons on the
assumption that the proton vertex is elastic. Exponential parameterisations
$e^{b t}$ seem to be appropriate.
The resulting distribution in slope parameter $b$
contains interesting features\cite{prn:cartiglia}. The $\xpom$
dependence of the slope parameter is consistent with a diffractive
interpretation at the smallest $\xpom$, with pion exchange becoming 
dominant for $0.1 \lapprox \xpom \lapprox 0.3$\cite{prn:pipred}.
For values of $\xpom$ above $0.3$, the
slope is best described by the soft colour interaction model.

\section{Exclusive Vector Meson Production}
\label{prn:vm}

Many results on vector meson production
were presented at the workshop, with
broad agreement between the different experiments on most points. The new 
developments in kinematic range are results at large $|t|$ for $\rho$, $\phi$
and $J/\psi$ photoproduction\cite{prn:adam}, $\rho$ 
electroproduction at low $Q^2$ 
in the HERA energy range\cite{prn:monteiro} and several
studies of vector meson production with proton 
dissociation\cite{prn:adam,prn:gaede}. In addition, new 
results from E665\cite{prn:e665} on $\rho$
production in the region $10 \lapprox W \lapprox 25 \ {\rm GeV}$ and
$0.15 < Q^2 < 20 \ {\rm GeV^2}$ were presented.

The photoproduction of light vector mesons continues to be well described by
soft pomeron exchange in conjunction with the vector dominance 
model\cite{prn:vdm}. The
$W$ dependence of $\rho$ photoproduction\cite{prn:gaede,prn:adam} matches 
parameterisations\cite{prn:ss} based on other soft physics data well. 
Results obtained by ZEUS for $\phi$\cite{prn:zeusphi} and 
$\omega$\cite{prn:zeusomega} photoproduction are also consistent with this
behaviour.
Shrinkage of the forward elastic peak in $\rho$ photoproduction
now seems to be established, with
ZEUS measuring $\alphapom^{\prime} = 0.23 \pm 0.15 \ ({\rm stat.}) 
^{+ 0.10}_{- 0.07} \ ({\rm syst.}) \ {\rm GeV^{-2}}$ from the difference in
slope parameters measured at HERA and at lower energy\cite{prn:adam}. This is
consistent with the figure established from $pp$ elastic 
scattering\cite{prn:cdfelas}. Shrinkage seems also to be present
in the low $Q^2$ electroproduction regime\cite{prn:schellman}.

A clear message from the workshop was that wherever a large scale 
($Q^2$, $t$ or a heavy quark mass) is 
introduced, the soft physics description breaks down, with perturbative
approaches being more appropriate. Under these circumstances, $W$ dependences
steepen, $t$-slope parameters fall and the skewing of the $\rho$ line-shape
diminishes. It is now clearly established from HERA data alone that the $W$
dependence of the $J/\psi$ photoproduction cross section is significantly
steeper than that predicted from soft pomeron 
models\cite{prn:gaede,prn:bella}.
Perturbative calculations\cite{prn:jpsipred} 
that predict the $W$ dependence of the cross 
section from the square of the proton
gluon density are in broad agreement with the data. It
is not yet established whether there is any shrinkage in the forward elastic
peak for $J/\psi$ photoproduction.

\begin{figure}[h] \unitlength 1mm
 \begin{center}
   \begin{picture}(136,85)
     \put(-1.7,42.5){\epsfig{file=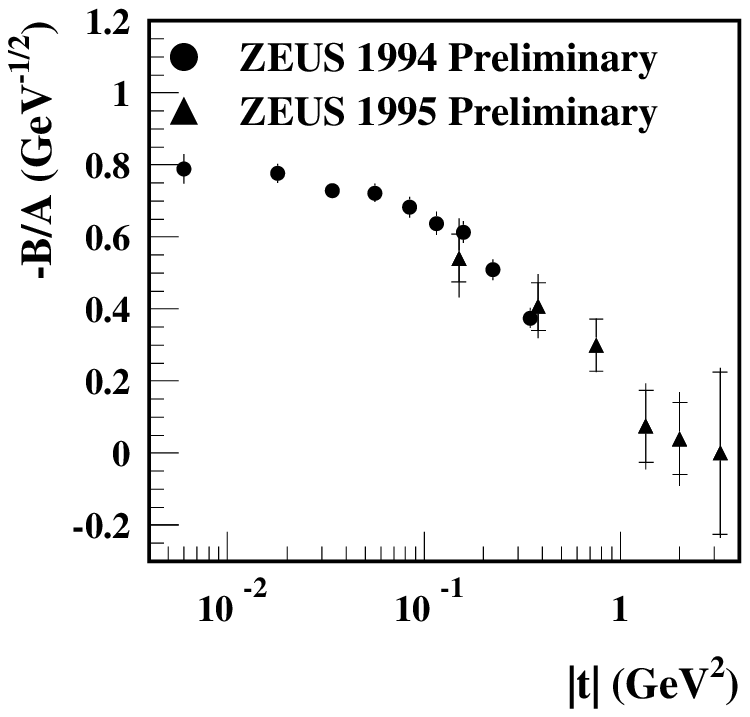,width=0.3825\textwidth}}
     \put(39.1,42.5){\epsfig{file=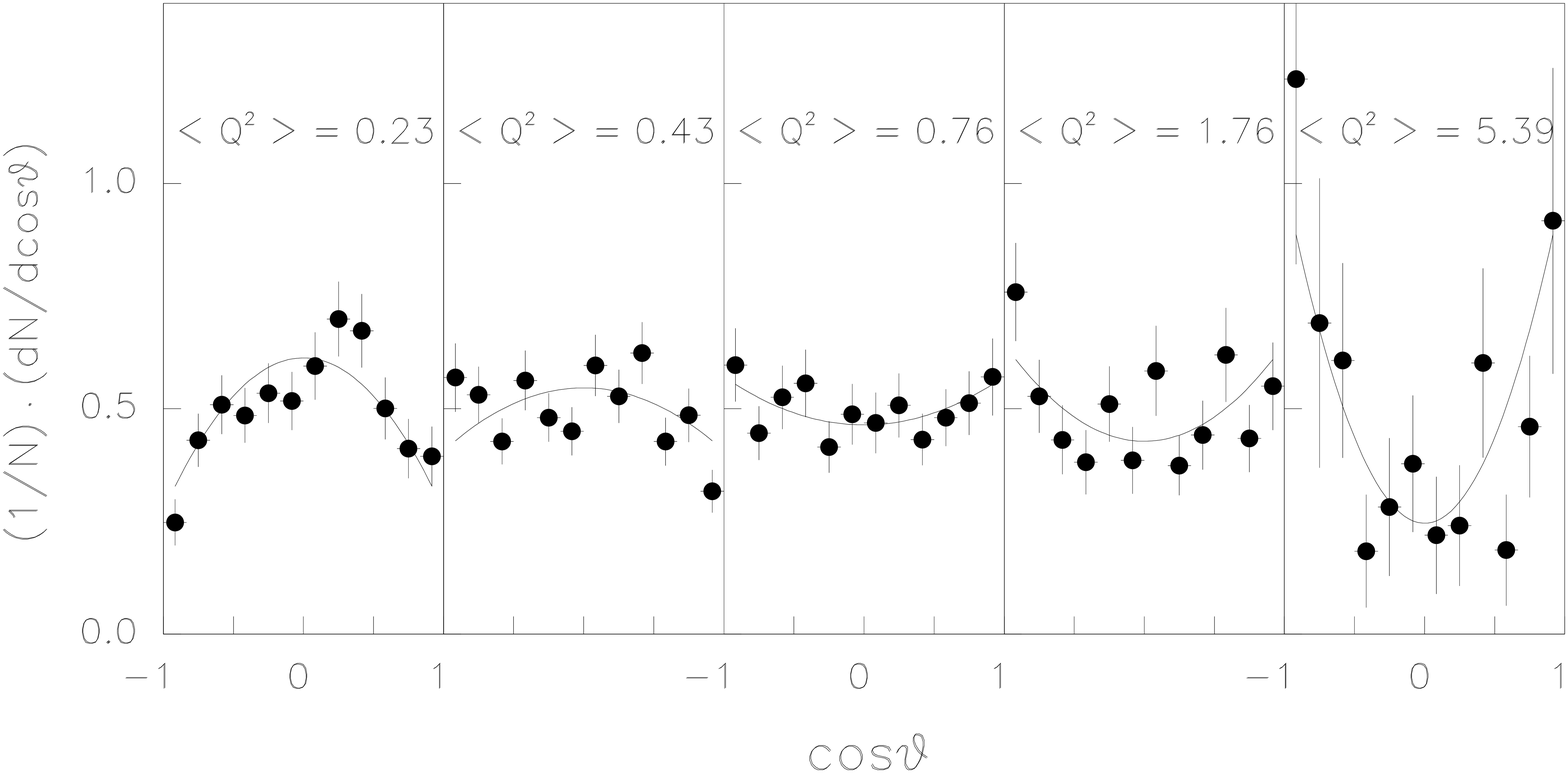,width=0.595\textwidth}}
     \put(1.7,0){\epsfig{file=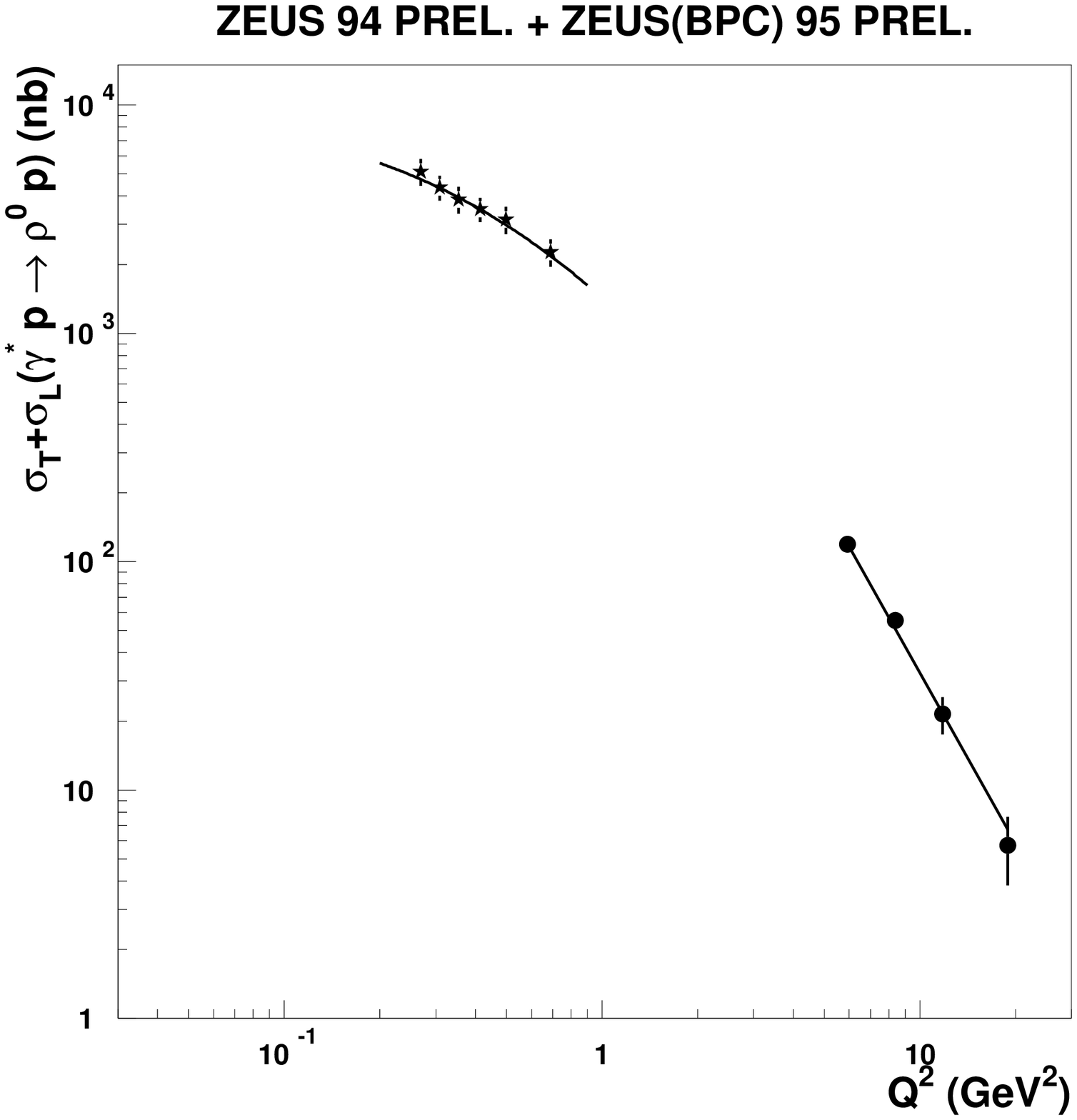,width=0.34\textwidth}}
     \put(51,-4.25){\epsfig{file=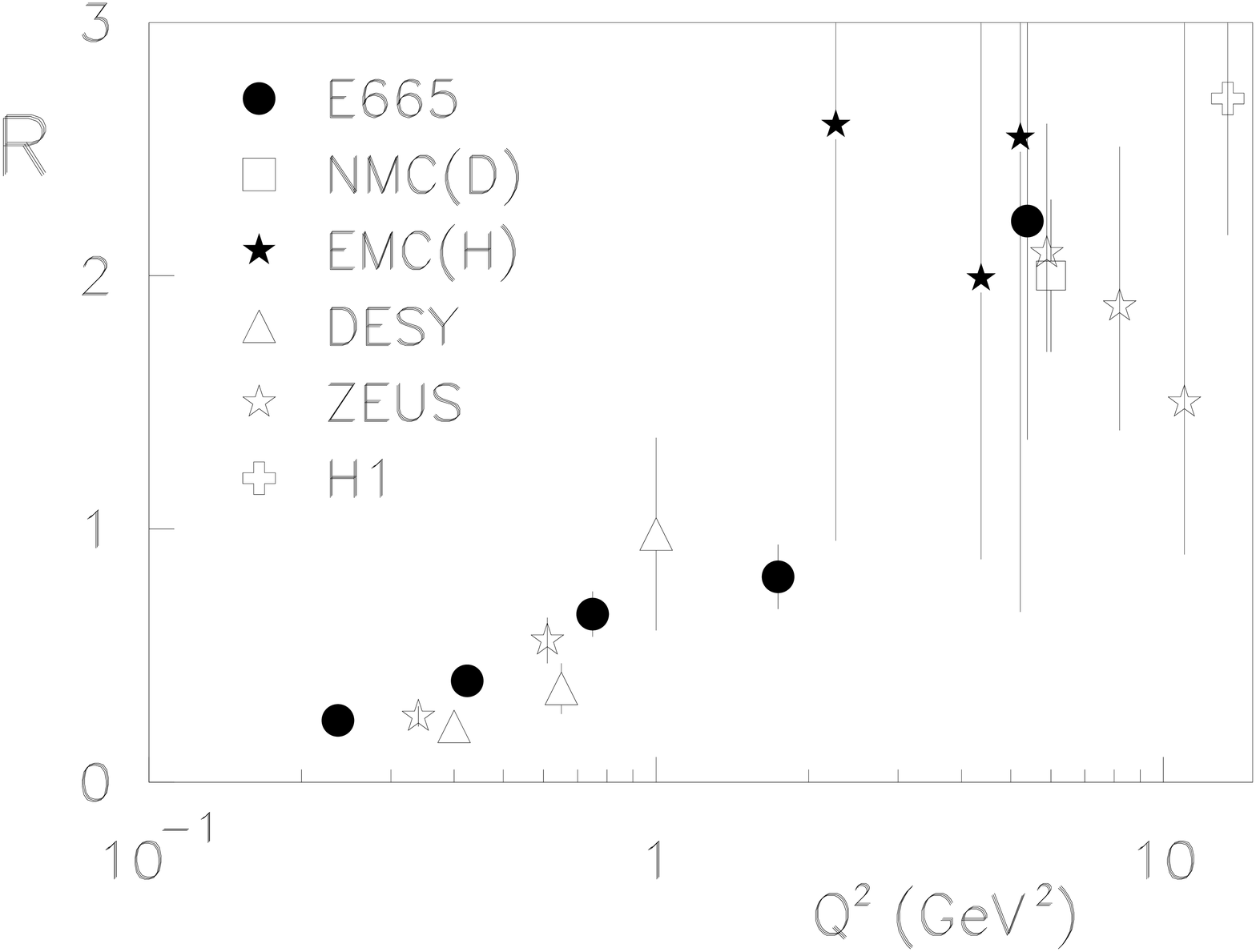,width=0.51\textwidth}}
     \put(23.8,36){\tiny{\bf $\av{W} = 52 \ {\rm GeV}$}}
     \put(20,17){\tiny{\bf $\av{W} = 56 \ {\rm GeV}$}}
     \put(16,39.95){\tiny{\bf $n = 1.94 \pm 0.12 \pm 0.18$}}
     \put(13,13){\tiny{\bf $n = 2.52 \pm 0.28 \pm 0.24$}}
     \put(12.75,59.5){\bf (a)}
     \put(8.5,8.5){\bf (b)}
     \put(56.1,59.5){\bf (c)}
     \put(86.7,34){\bf (d)}
   \end{picture}
 \end{center}
 \vspace{0.2cm}
 \scaption {Scale dependence of various vector meson parameters. (a) The 
ratio $-B / A$ in S\"{o}ding fits to the $\rho$ line-shape in photoproduction,
giving a measure of the degree of skewing
as a function of $t$\protect \cite{prn:adam} \protect. (b) The $Q^2$ 
dependence of the total elastic $\rho$ cross 
section\protect \cite{prn:monteiro} \protect together with the results for
$n$ of fits to equation~\ref{prn:vdm}. (c) Decay angular
distributions for $\rho$ electroproduction as a function of 
$Q^2$\protect \cite{prn:schellman} \protect, clearly showing the transition
from transverse to longitudinal polarisation. (d) The $Q^2$ 
dependence of the ratio $R$ of
longitudinal to transverse photon $\rho$ production cross 
sections\protect \cite{prn:schellman} \protect.}
 \label{prn:vmfigs}
\end{figure}

The skewing of the $\rho^0$ line-shape, usually interpreted as the result of
interference between resonant and non-resonant di-pion 
production\cite{prn:soeding} becomes less significant with increasing 
$t$\cite{prn:adam} (see figure~\ref{prn:vmfigs}a) 
as well as $Q^2$\cite{prn:monteiro}. The $\phi:\rho$
and $J/\psi:\rho$ ratios also increase with both of these 
scales\cite{prn:adam,prn:gaede}.

As $Q^2$ is increased, the $W$ dependence of $\rho$ production 
appears to become steeper when HERA measurements are compared with NMC 
data\cite{prn:nmc}. However, the new results from E665\cite{prn:e665} show a
larger cross section at low energy, yielding a $W$ dependence that is still
compatible with the soft pomeron.
The difference between the low energy measurements is likely to be at
least in part related to the assumptions made regarding proton dissociation
background\cite{prn:schellman}. With the present experimental precision, it is
not possible to reslolve this question using HERA data alone.
The $t$-slope parameter for $\rho$ production clearly falls 
with increasing $Q^2$ at fixed $W$\cite{prn:monteiro}, as expected from
the decrease in transverse separation of $q \bar{q}$ components of the photon
with increasing $Q^2$.
There is also a clear decrease in slope parameters when the proton-dissociative
$\rho$ production process is compared with its elastic counterpart, both in 
photoproduction\cite{prn:adam} and at high 
$Q^2$\cite{prn:gaede,prn:h1pdis}.
This reflects a similar behaviour in $pp$ interactions\cite{prn:reviews}. 
For the $J/\psi$,
the slope parameter is already significantly smaller than that for the $\rho$
in the photoproduction regime. Results on $J/\psi$ 
electroproduction\cite{prn:bella} do not
indicate any differences in the $W$ or $t$ dependences compared to those
at $Q^2 = 0$.

The $Q^2$ dependence of the total cross section for
vector meson production is fitted to the form
\begin{eqnarray}
\sigma (Q^2)  = \sigma (Q^2 = 0) 
\left( \frac{m_{_{\rm V}}^2}{Q^2 + m_{_{\rm V}}^2} \right)^n \ ,
\label{prn:vdm}
\end{eqnarray} 
where $m_{_{\rm V}}$ is the vector meson mass. In the vector dominance model
$n \sim 2$ is expected\cite{prn:bauer}. The
results for both the $\rho$ (see figure~\ref{prn:vmfigs}b) and the 
$J/\psi$\cite{prn:schellman,prn:monteiro,prn:gaede} are found to lie in the
region $n = 2.0 - 2.5$.

The $Q^2$ dependence of the vector meson polarisation has now been extensively
studied, through the distributions in the angle $\theta^{\star}$ in the rest
frame of the $\rho$ between the direction of the positively charged decay
pion and the $\rho$ direction in the $\gamma^{\star} p$ centre of mass frame.
In photoproduction, both the $\rho$ and $J/\psi$ are consistent
with full transverse polarisation\cite{prn:gaede,prn:bella}, as expected
from $s$-channel helicity conservation. At large $Q^2$, the transition to
a dominantly longitudinal vector meson polarisation is rather 
rapid, as can be seen from the change in the $\cos \theta^{\star}$
distribution as measured by E665 
(figure~\ref{prn:vmfigs}c)\cite{prn:schellman}. 
Natural parity exchange and $s$-channel
helicity conservation have been explicitly demonstrated from the large $Q^2$
data\cite{prn:schellman,prn:monteiro} using the methods of\cite{prn:spin},
such that the longitudinal to transverse photon
cross section ratio $R (Q^2)$ can be
extracted. A compilation of measurements of this ratio is shown in 
figure~\ref{prn:vmfigs}d.

A final area of study is radial excitations of vector mesons. The 
$\psi (2S) : J/\psi$ photoproduction ratio is measured by H1 to be
$0.16 \pm 0.06$\cite{prn:gaede}, consistent with predictions based
on the convolution of the $1S$ and $2S$ wavefunctions with the spatial 
separation of the diquarks in the process 
$\gamma^{\star} \rightarrow q \bar{q}$\cite{prn:kop}. 
The $\rho^{\prime} : \rho$
ratio is larger at high $Q^2$ than in photoproduction\cite{prn:gaede}, as
predicted in the same model.

\section{Summary}

Significant developments in colour-singlet exchange physics have taken 
place in the past year, both from an experimental and a phenomenological
point of view. 
There is now agreement that the value of the
intercept of the diffractive trajectory dominating the photon dissociation
process
at high $Q^2$ is larger than that extracted in photoproduction or from soft
hadron-hadron data.
There is a consensus, arising both from inclusive measurements and from
final state analyses, that the exchange mediating the
diffractive process in DIS invloves a gluon carrying a large fraction 
of the momentum at low scales.
Comparisons of parton distributions extracted
from diffractive structure function analyses with dijet production rates both
at HERA and the Tevatron have begun. There are already indications that 
incomplete rapidity gap survival probabilities may play an important role.
Cross section measurements for leading baryon production at large $\xpom$ are
developing fast and their description presents a considerable challenge to 
theorists. The new vector meson data map out the transition
from the vector dominance to the perturbative region with increasing precision.
\vspace{-0.2cm}
\section*{Acknowledgments}

It is a pleasure to thank the speakers in the diffractive sessions for an
excellent set of talks and the other conveners, Dave Soper, Amadeo Staiano and
Phillip Melese, with whom I enjoyed working very much. Due to the
efforts of Jose Repond and his team, the workshop was extremely well
organised. I am grateful to John Dainton and Rosario Nania for proof reading
this manuscript.

\end{document}